\documentclass[aip,cha,amsmath,amssymb,preprint,floatfix]{revtex4-1}

\usepackage{graphicx}
\usepackage{dcolumn}
\usepackage{bm}

\makeatother

\begin{document}

\title{Riddling: chimera's dilemma}

\author{V. Santos}
\affiliation{Graduate in Science Program, State University of Ponta
Grossa, Ponta Grossa, Paran\'a, 84030-900, Brazil.}
\author{J. D. Szezech Jr}
\affiliation{Graduate in Science Program, State University of Ponta
Grossa, Ponta Grossa, Paran\'a, 84030-900, Brazil.}
\affiliation{Department of Mathematics and Statistics, State University of
Ponta Grossa, Ponta Grossa, Paran\'a, 84030-900, Brazil.}
\email{jdsjunior@uepg.br}
\author{A. M. Batista}
\affiliation{Graduate in Science Program, State University of Ponta
Grossa, Ponta Grossa, Paran\'a, 84030-900, Brazil.}
\affiliation{Department of Mathematics and Statistics, State University of
Ponta Grossa, Ponta Grossa, Paran\'a, 84030-900, Brazil.}
\affiliation{Potsdam Institute for Climate Impact Research, Potsdam,
Brandenburg, 14473, Germany.}
\author{K. C. Iarosz}
\affiliation{Potsdam Institute for Climate Impact Research, Potsdam,
Brandenburg, 14473, Germany.}
\affiliation{Institute of Physics, University of S\~ao Paulo, S\~ao Paulo,
S\~ao Paulo, 05508-900, Brazil.}
\affiliation{Department of Physics, Humboldt University, Berlin, Brandenburg,
12489, Germany.}
\author{M. S. Baptista}
\affiliation{Institute for Complex Systems and Mathematical Biology, SUPA,
University of Aberdeen, Aberdeen, AB24 3UE, Scotland, United Kingdom.}
\author{H. P. Ren}
\affiliation{Shaanxi Key Laboratory of Complex System Control and Intelligent
Information Processing, Xian University of Technology, Xi'an, 710048, PR China.}
\affiliation{Xian Technological University, Xi'an, 710021, PR China.}
\author{C. Grebogi}
\affiliation{Institute for Complex Systems and Mathematical Biology, SUPA,
University of Aberdeen, Aberdeen, AB24 3UE, Scotland, United Kingdom.}
\affiliation{Shaanxi Key Laboratory of Complex System Control and Intelligent
Information Processing, Xian University of Technology, Xi'an, 710048, PR China.}
\author{R. L. Viana}
\affiliation{Department of Physics, Federal University of Paran\'a, Curitiba,
Paran\'a, 80060-000, Brazil.}
\author{I. L. Caldas}
\affiliation{Institute of Physics, University of S\~ao Paulo, S\~ao Paulo,
S\~ao Paulo, 05508-900, Brazil.}
\author{Y. L. Maistrenko}
\affiliation{Potsdam Institute for Climate Impact Research, Potsdam,
Brandenburg, 14473, Germany.}
\affiliation{Institute of Mathematics and Centre for Medical and Biotechnical
Research, National Academy of Sciences of Ukraine, Tereshchenkivska St. 3,
01030 Kyiv, Ukraine.}  
\author{J. Kurths}
\affiliation{Potsdam Institute for Climate Impact Research, Potsdam,
Brandenburg, 14473, Germany.}
\affiliation{Department of Physics, Humboldt University, Berlin, Brandenburg,
12489, Germany.}

\date{6 August 2018}

\begin{abstract}
We investigate the basin of attraction properties and its boundaries for
chimera states in a circulant network of H\'enon maps. It is known that
coexisting basins of attraction lead to a hysteretic behaviour in the diagrams
of the density of states as a function of a varying parameter. Chimera states,
for which coherent and incoherent domains occur simultaneously, emerge as a
consequence of the coexistence of basin of attractions for each state.
Consequently, the distribution of chimera states can remain invariant by a
parameter change, as well as it can suffer subtle changes when one of the
basins ceases to exist. A similar phenomenon is observed when perturbations are
applied in the initial conditions. By means of the uncertainty exponent, we
characterise the basin boundaries between the coherent and chimera states, and
between the incoherent and chimera states, respectively. This way, we show that
the density of chimera states can be not only moderately sensitive but also
highly sensitive to initial conditions. This chimera's dilemma is a consequence
of the fractal and riddled nature of the basins boundaries. 
\end{abstract}

\maketitle

\begin{quotation}
Coupled dynamical systems have been used to describe the behaviour of real
complex systems, such as power grids, neuronal networks, economics, and chemical
reactions. Furthermore, these systems can exhibit various kinds of interesting
nonlinear dynamics, e.g. synchronisation, chaotic oscillations, and chimera
states. The chimera state is a spatio-temporal pattern characterised by the
coexistence of coherent and incoherent dynamics. It has been observed in a
great variety of systems, ranging from theoretical and experimental arrays of
oscillators, to in phenomena such as the unihemispheric sleep of cetaceans. We
study the chimera state in a circulant network of H\'enon maps, seeking to
determine how the density of states in the network depends on the system
parameters and the initial conditions. We have found that, as expected, the
density of states might be invariant to parameter alterations, but it might
also tip when a basin of attraction ceases to exist. When the basin boundary of
the chimera states is fractal, the densities of the states will depend
moderately on the perturbations in the initial conditions, and they may even
remain invariant to specific perturbations. However, when the basin boundary
is riddled, even arbitrarily small perturbations to the initial conditions can
replace the chimera state to an incoherent state. The existence of basin
boundary in a network that presents chimera states is a chimera's dilemma. 
\end{quotation}

\section{INTRODUCTION}

Chimera state, in reference to the Greek mythological creature, is a
spatio-temporal pattern observed in coupled dynamical systems that was first
reported by Kuramoto and Battogtokh in 2002 \cite{kuramoto02}. This pattern is
characterised by the coexistence of coherent and incoherent behaviours
\cite{umberger89,abrams04,abrams08,omelchenko11,dudkowski14}. It has been
identified in paradigmatic network models \cite{omelchenko08,andrzejak17}, such
as the Kuramoto model \cite{santos15,yao15}, networks of Hindmarsh-Rose neurons
\cite{santos17}, and coupled van der Pol-Duffing oscillators
\cite{dudkowski16}. Chimera states have also been found in experimental
se\-ttings \cite{tinsley12}. Martens et al. \cite{martens13} showed them in a
mechanical experiment composed of coupled metronomes. Kapitaniak et al.
\cite{kapitaniak14} demonstrated the formation of chimera in Huygens's clocks
realised by metronomes. Coupled electronic oscillators can exhibit chimera with
quiescent and synchronous domains \cite{gambuzza14}. 

Basins of attraction for chimera states were analysed by Martens et al.
\cite{martens16}. They considered two coupled populations of Kuramoto-Sakaguchi.
The chimera states have a coexisting asynchronous and synchronous population,
where their basins of attraction show a complex twist structure. Rakshit et al.
\cite{rakshit17} identified and quantified incoherent, coherent, and chimera
states in coupled time-delayed Mackey-Glass oscillators by means of basin
stability analysis. The coexisting basins were found to be roughly robust to
the coupling strength and coupling radius alterations in certain network
configurations, i.e. the density of the chimera states could be preserved by
the coupling strength and the coupling radius alterations for those
configurations. Our interest is to understand this stability of the density of
the states in terms of initial conditions. To this goal, we analyse a
circulant network composed of H\'enon maps and characterise its basin
boundaries for chimera states.

The H\'enon map was proposed as a simplified model to study the dynamics of the
Lorenz model \cite{henon69}. Networks of coupled H\'enon maps have been
considered in studies about periodic orbits \cite{politi92}, chaotic dynamics
of spatially extended systems \cite{astakhov01}, and unstable dimension
variability structure \cite{santos16}. Semenova et al. have recently found
chimera states in ensembles of non-locally coupled H\'enon maps
\cite{semenova17}. They also explored the effects of noise perturbations on the
network.

In this work, we calculate the strength of incoherence to identify incoherent,
coherent, and chimera states. Clearly, each network state (coherent or
incoherent) has its own basin of attraction. Parameter changes modify the
Lebesgue measure of the basins, which in extreme situations can cease to exist,
leaving a network whose nodes will be either in the coherent or incoherent
states. Our main interest, however, is to understand how perturbations in the
initial conditions change the density of these states in the network. To this
goal, we study the property of the basins of attraction's boundaries. We find
that whereas the basin boundary between the incoherent and chimera state are
typically riddled, the basins boundary between the chimera and the coherent
state is typically fractal. Thus, small alterations in the initial conditions
can always change the density of the states. However, arbitrarily small
perturbations in the initial conditions can shift a chimera state to an
incoherent one.

Riddled basin is a basin of attraction (of an attractor) such that every point
of it has pieces of another attractor's basin arbitrarily nearby
\cite{alexander92,ott93,ashwin94}. A riddled basin of attraction has the same
fractal dimension of its boundary. Heagy et al. \cite{heagy94} reported
experimental and numerical evidence of riddled basins in coupled chaotic
systems. They studied chaos synchronisation in coupled chaotic oscillator
circuits. Woltering and Markus \cite{woltering00} identified the existence of
riddled basin in a model for the Belousov-Zabotinsky reaction.

This paper is organised as follows: Section 2 introduces the network of coupled
maps. In Section 3, we present the basin of attraction for chimera states and
our results for the basin boundaries. In the last Section, we draw our
conclusions.


\section{NETWORK MODEL}

\begin{figure}[hbt]
\centering
\includegraphics[width=0.9\linewidth]{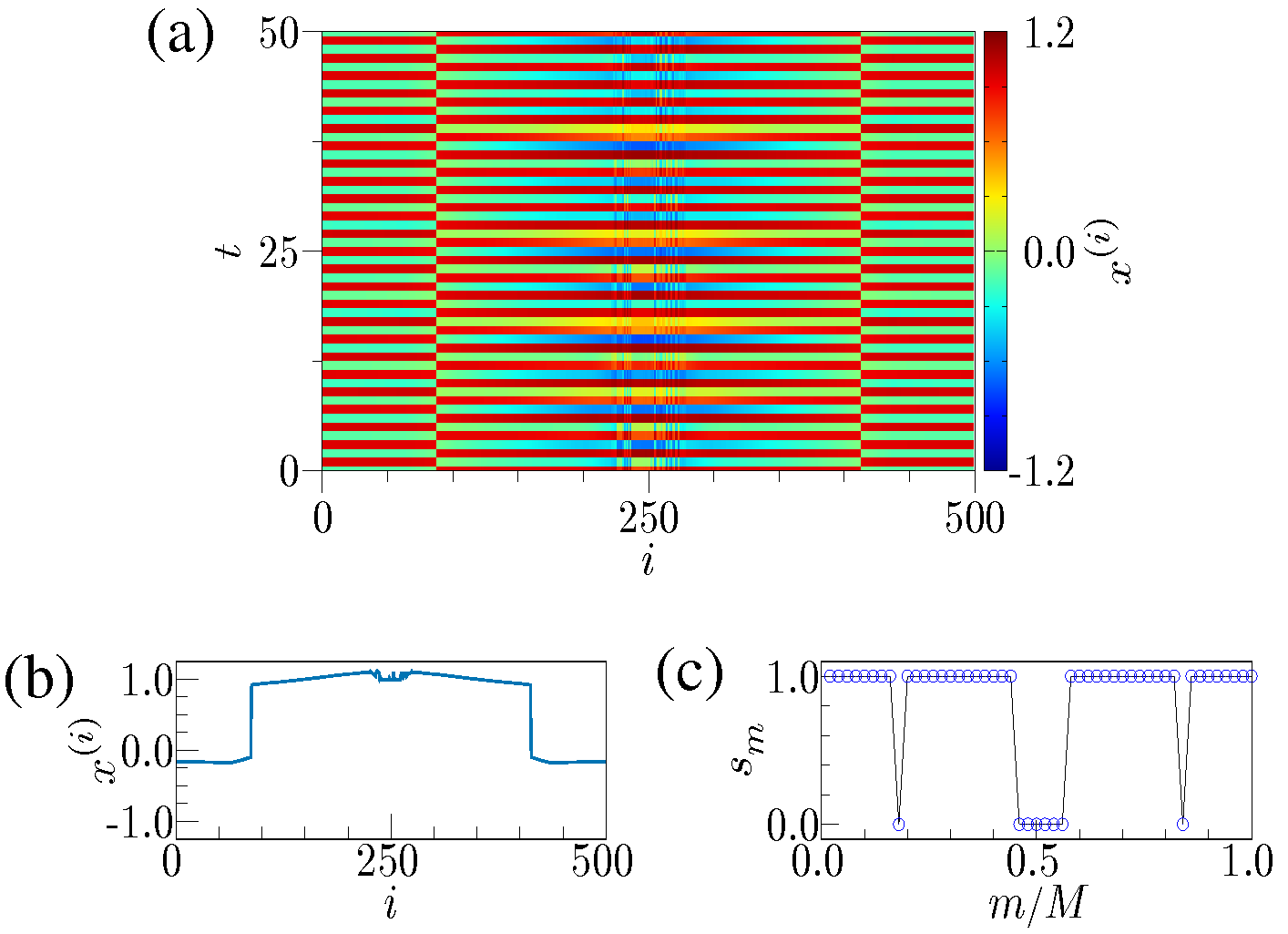}
\caption{(Colour online) (a) Space-time plot of the dynamics of the network
Eq. \eqref{eq:MapCoup} after the transient time, where the colour bar gives the
value of the $x$ variable of each map in the network. In (b) and (c) we plot
the snapshot and its $s_m$ spectrum, respectively, for $t=26$ of (a). We
consider $\alpha=1.44$, $\beta=0.164$, $\sigma=0.30$, and $r=0.30$.}
\label{fig1}
\end{figure}

Networks of coupled maps have been used to study extended dynamical system
\cite{kaneko92}. We consider a network composed of $N$ coupled H\'enon maps
written as
\begin{equation}\label{eq:MapCoup}
\mathbf{x}_{t+1}^{(i)}=\mathbf{F}(\mathbf{x}_t^{(i)})+
\frac{\sigma \mathbf{E}}{2rN}\sum_{j=i-rN}^{i+rN}[\mathbf{F}(\mathbf{x}_t^{(j)})-
\mathbf{F}(\mathbf{x}_t^{(i)})],
\end{equation}
where $i=1,\ldots ,N$, $t$ is the discrete time,
$\mathbf{F}(\mathbf{x})=[1-\alpha x^2+y,\beta x]^T$ is the two-dimensional
H\' enon map, $\sigma$ and $r$ are the coupling intensity and coupling radius,
respectively, and
\begin{equation}
\mathbf{E}=
\begin{pmatrix}
1 & 0 \\
0 & 0
\end{pmatrix},
\end{equation}
specifies which variables of the H\'enon map are coupled here, namely $x$. This
system was previously studied by Semenova et al. \cite{semenova15} for the
parameter set $(\alpha,\beta)=(1.4,0.3)$ focusing on the parameter space
$\sigma\times r$. In our network, we use $(\alpha,\beta)=(1.44,0.164)$, because
the H\' enon map exhibits a period-$5$ attractor for these parameters.
We consider a circulant network of H\'enon maps. Figure \ref{fig1}(a) shows
the spatio-temporal plot obtained from Eq. \eqref{eq:MapCoup} for $\sigma=0.30$
and $r=0.30$, where the colour bar represents the variable $x^{(i)}$. We find
two coherent and one incoherent (small region around $i=250$) domains, as shown
in Fig. \ref{fig1}(b). The discontinuities in $x^{(87)}$ and $x^{(412)}$ are due
to the splitting of the spatial profile into two branches, while the interval
region from approximately $x^{(220)}$ to $x^{(280)}$ displays spatial incoherence
(irregular spatial pattern). A chimera state of the form as in Figs.
\ref{fig1}(a) and \ref{fig1}(b) was first reported by Omelchenko et al.
\cite{omelchenko12}.

Aiming to characterise coherent and incoherent states, we use a quantitative
measure proposed by Gopal et al. \cite{gopal14}. To do that, first, we
calculate $s_m=\Theta(\delta-\chi(m))$, where $\Theta$ is the Heaviside step
function and $\delta$ is a predetermined threshold. The local standard
deviation $\chi^{(l)}(m)$ is given by 
\begin{equation}\label{eq:sdevm_def}
\chi^{(l)}(m)=\left\langle\sqrt{\frac{1}{n}\sum^{nm}_{j=n(m-1)+1}\left[ z^{(l,j)}-
\langle z^{(l)}\rangle\right]^2}\right\rangle_t,
\end{equation}
where $n=N/M$, $m=1,2,\dots,M$ and
$\mathbf{z}_t^{(i)}=\mathbf{x}_t^{(i)}- \mathbf{x}_t^{(i+1)}$ with
$\mathbf{z}^{(i)}=[z^{(1,i)},z^{(2,i)},\dots,z^{(d,i)}]^T\in\mathbb{R}^d$, and $d$ is 
the dimension of the dynamical system. In these new variables, two neighbouring
oscillators describing a node of the network $i$ and $i+1$ are oscillating
coherently if $\mathbf{z}^{(i)}\approx 0$, and incoherently otherwise. 
$\left\langle z^{(l)}\right\rangle=\frac{1}{n}\sum^{nm}_{j=n(m-1)+1}z_t^{(l,j)}$ is
the average of $z_t^{(l)}$ over the partition $m$ for a fixed time, and
$\left\langle\dots\right\rangle_t$ is the time average. We set $\delta=1\%$ of
$|x^{(l,max)}-x^{(l,min)}|$, and the network size $N=500$. Figure \ref{fig1}(c)
shows $s_m$ for the network separated into $M=50$ partitions. By means of $s_m$
versus $m/M$, we can clearly identify the coherent and incoherent regions.

\begin{figure}[hbt]
\centering
\includegraphics[width=0.9\linewidth]{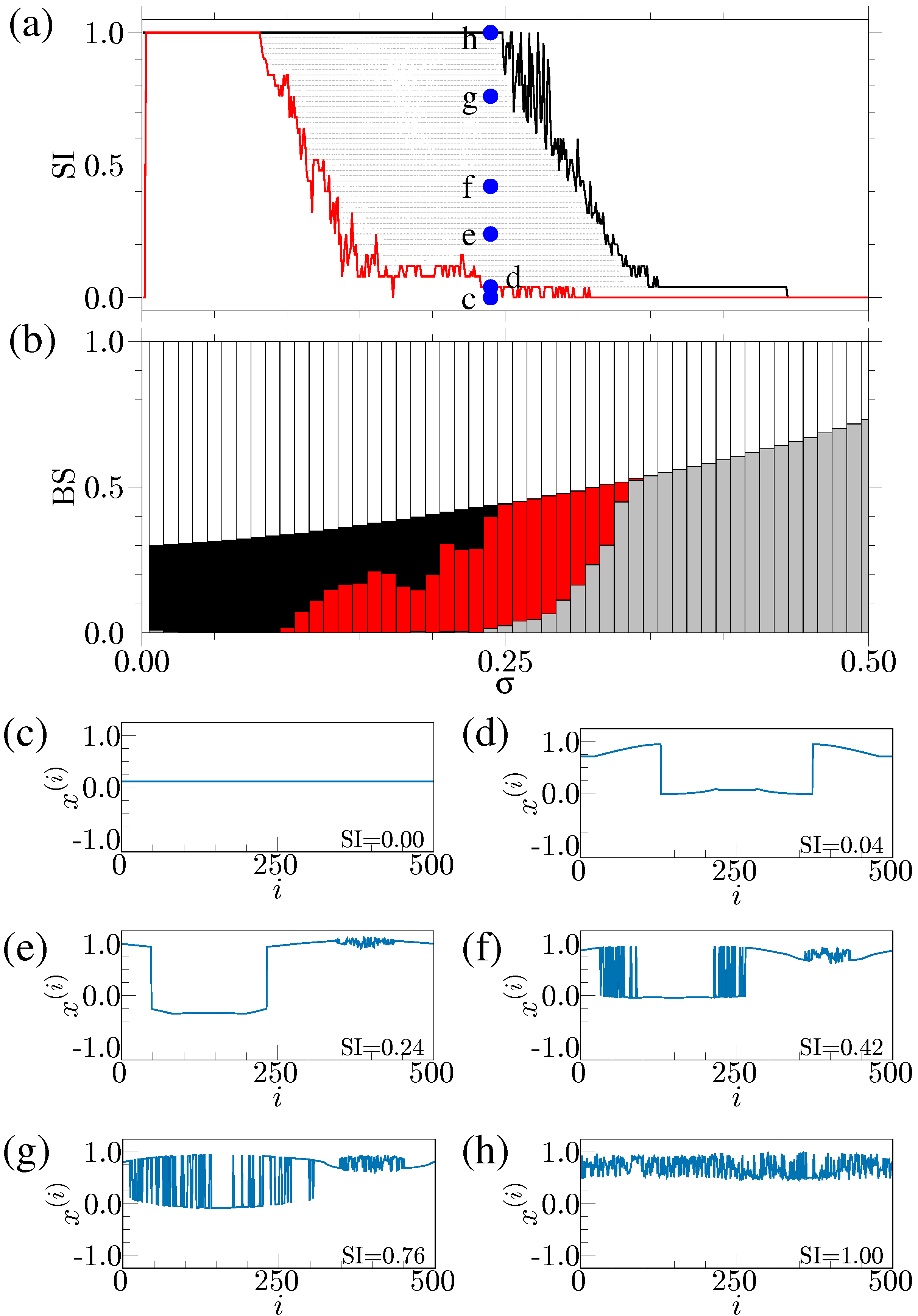}
\caption{(Colour online) (a) SI versus $\sigma$ for $400$ different initial
conditions. The red (black) line outlines the minimum (maximum) value of SI.
(b) BS versus $\sigma$ for incoherent (black), chimera (red), coherent
(gray) states, and divergent (vertically dashed). From (c) to (h) we plot some
coexistent states for $\sigma=0.24$. We consider $\alpha=1.44$, $\beta=0.164$,
and $r=0.30$.}
\label{fig2}
\end{figure}

Gopal et al. \cite{gopal14} developed the measure {\it strength of incoherence}
(SI) to characterise the spatial dynamics of nonlinear coupled networks. It is
able to identify coherent and incoherent states, as well as chimera states
\cite{rakshit17,gopal14}. The SI is given by
\begin{equation}\label{eq:SI_def}
{\rm SI}=1-\frac{\sum^{M}_{m=1}s_m}{M}.
\end{equation}
If $\chi^{(l)}(m)>\delta$, some of the oscillators in the $m\textrm{-th}$
partition are incoherent and $s_m=0$. When
$N\rightarrow\infty$, ${\rm SI}\rightarrow 1$ ($s_m=0,\forall m$) for
incoherent states, ${\rm SI}\rightarrow 0$ for coherent and cluster states, and
$0<{\rm SI}<1$ for chimera states. In Fig. \ref{fig2}(a), we plot SI versus the
coupling strength $\sigma$ for $400$ different initial conditions of the system
(\ref{eq:MapCoup}). We consider $(x_0^{(i)},y_0^{(i)})=(0,0)$ for $i=2,\cdots,N$
and $(x_0^{(1)},y_0^{(1)})$ is homogeneously distributed in the interval
$[-3,3]\times[-3,3]$. The state variable is iterated $10,500$ times, with the
first $9,000$ being discarded as transient state, and the last $1,500$ are
included to calculate SI. The accuracy of our results is not improved by
doubling the size of the dataset. The long transient is considered to avoid
treating transient chimera states as an asymptotic state.

Figure \ref{fig2}(a) shows the coexistence of multiple sta\-tes with different
values of SI for the same $\sigma$ in the interval $[0.08,0.44]$. This
hysteresis course reflects that the basin of attraction for the coherent and
the incoherent states coexist. For smaller values of the coupling strength
there is only the incoherent state (characterised by the red curve for
${\rm SI}=1$) and its large basin of attraction occupies a large domain of
initial conditions considered (excluding the infinity basin). About
$\sigma\approxeq 0.08$, the coexistence of three basins of attractions causes
the network to behave either in the coherent state (smaller SI values), the
incoherent state (larger SI values), or in the chimera (intermediate SI
values). Appropriately chosen initial conditions may lead a network whose
$\sigma$ is being altered to have states characterised by the red curve until
$\sigma=0.5$. For intermediate $\sigma$ values, the network is characterised by
coherent and chimera states with lower SI values. At $\sigma=0.5$, there is
only the basin of attraction for the coherent states. For appropriately chosen
initial conditions, as $\sigma$ is varied from $0.5$ to zero, the network might
present a distinct route from coherence to incoherence (characterised by the SI
for the black curve). This means that the network has a hysteresis behaviour
for its states, typical to happen in networks which present chimera. Figure
\ref{fig2}(b) exhibits the single node basin stability (BS) as a function of
$\sigma$ for incoherent (black), chimera (red), coherent (gray), and divergent
(white) states. BS is associated with the volume of the basin of attraction
\cite{menck13,menck14,schultz17}. In Figs. \ref{fig2}(c)-(h) we plot snapshots
of the dynamic behaviour for $\sigma=0.24$. Changing the initial conditions of
only one map of the network, we observe: (c) synchronised period-$5$ dynamics
corresponding to ${\rm SI}=0.00$, (d) period-$2$ cluster state in which
${\rm SI}=0.04$, (e) to (g) chimera states for different sizes of incoherent
states with ${\rm SI}=0.24$, ${\rm SI}=0.42$, and ${\rm SI}=0.76$,
respectively, and (h) incoherent state for which ${\rm SI}=1.00$.


\section{BASIN OF ATTRACTION FOR CHIMERA STATES}

In our network, for some values of $\sigma$ a great variety of dynamical states
can be found by only changing the initial conditions. With this in mind, we
investigate this phenomenon by means of the basin of attraction. To do that, we
construct a grid and vary the initial values of the variables of one map of the
network, while the others are kept equal to $0$. 

\begin{figure}[hbt]
\centering
\includegraphics[width=0.6\linewidth]{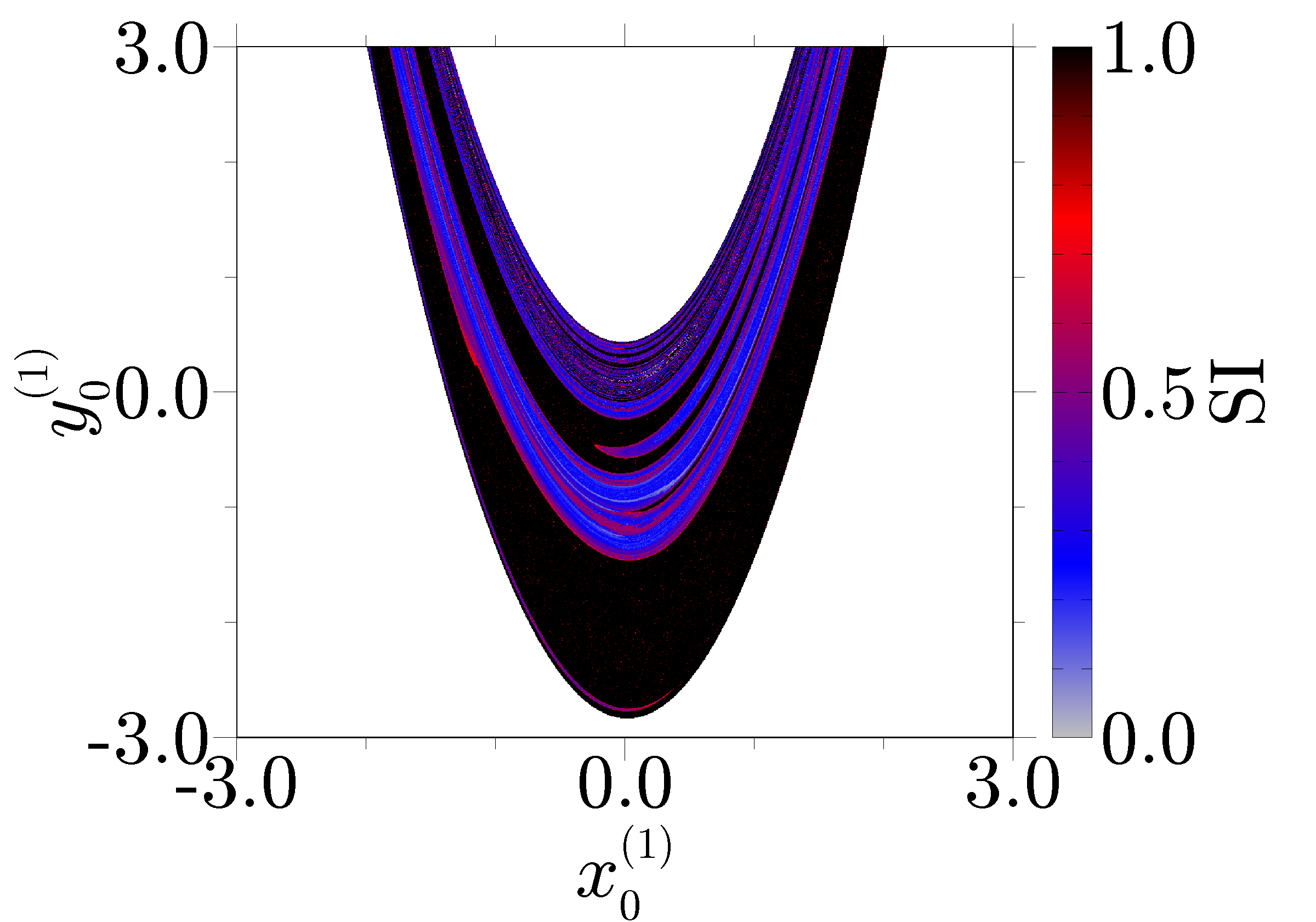}
\caption{(Colour online) Basin of attraction of only $1$ H\'enon map in the
network with $\sigma=0.18$, $\alpha=1.44$, $\beta=0.164$, and $r=0.30$, where
the colour bar represents the SI values. The black points correspond to
incoherent states, the grey points denote the synchronised cluster states, from
blue to red points represent the chimera states. The initial conditions in the
white region diverge to infinity.}
\label{fig3}
\end{figure}

In Fig. \ref{fig3}, we plot the basin of attraction for $\sigma=0.18$ with the
SI values being represented by a colour scale. It displays the same overall
shape of the basin of one individual H\'enon map. From Fig. \ref{fig3} it can
be noted that the density of each state varies depending on the region where we
sort the initial conditions, also in some regions the boundaries between the
basins may be very complex.

\begin{figure}[hbt]
\centering
\includegraphics[width=0.9\linewidth]{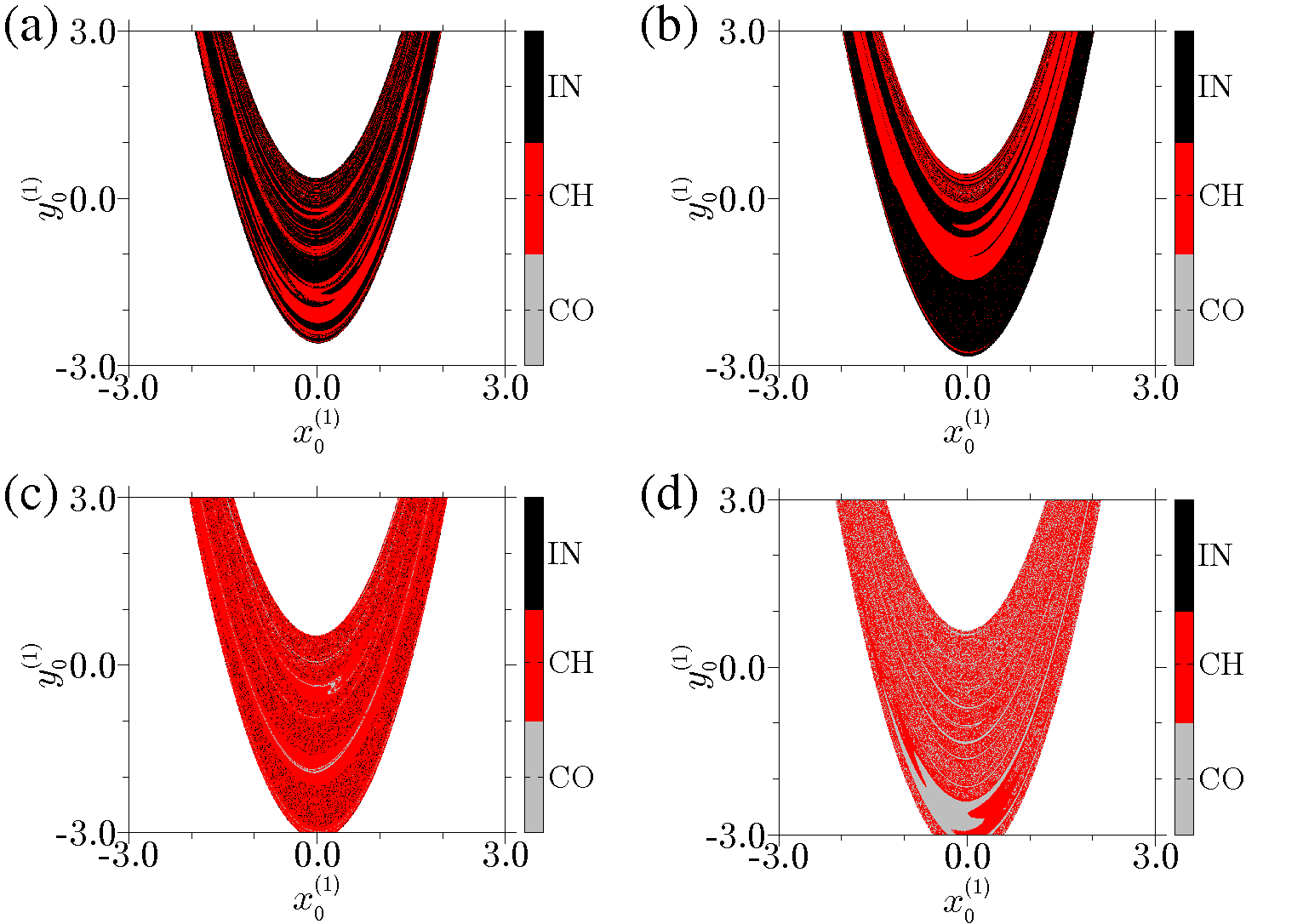}
\caption{(Colour online) Basins of attraction of the network of coupled H\'enon
maps for coherent (CO), chimera (CH), and incoherent (IN) states. We consider
$\sigma$ equal to (a) $0.12$, (b) $0.18$, (c) $0.24$, and (d) $0.30$.}
\label{fig4}
\end{figure}

In order to analyse the basin boundaries, we define ${\rm SI}\leq 0.04$ as
coherent state, ${\rm SI}\geq 0.90$ as incoherent state, and intermediate
values as chimera states. Applying these thresholds, we plot the basin for
$\sigma=0.12,0.18,0.24$, and $0.30$, as shown in Fig. \ref{fig4}, with gray
standing for coherent (CO), red for chimera (CH), and black for incoherent (IN)
states. When $\sigma$ is small, there is a predominance of incoherent and
chimera states in the basins. Increasing the value of $\sigma$, we find a
decrease in the size of the basin for incoherent states and an increase in that
for coherent states. The basins are arranged in a complicated way with some
regions exhibiting an apparent fractal structure. It was demonstrated that
fractality in the basin boundary can strongly affect the predictability of
final states in dynamical systems \cite{mcdonald85}.

The characterisation of basin boundaries can be made using the initial
condition uncertainty fraction, as introduced by McDonald et al.
\cite{mcdonald85}. The method consists of calculating the final state of a
number $N_0$ of random initial conditions in a region of the basin. If the
final state from a point in the center of a neighbourhood of radius
$\varepsilon$ is different from at least one of its neighbours, then such an
initial condition is $\varepsilon$-uncertain. The fraction of uncertain points
$f(\varepsilon)$ as a function of $\varepsilon$, for small $\varepsilon$, is
expected to scale according to $f(\varepsilon)\sim\varepsilon^\gamma$, where
$\gamma$ is the uncertainty exponent \cite{grebogi87,aguirre09}. The $\gamma$ is
related to the boundary of the sets being considered (in here they are the
basin of attractions) by $d=D-\gamma$, where $d$ is the dimension of the basin
boundary and $D=2$ is the phase space dimension of the boxes used to calculate
$\gamma$. 

\begin{figure}[hbt]
\centering
\includegraphics[width=0.9\linewidth]{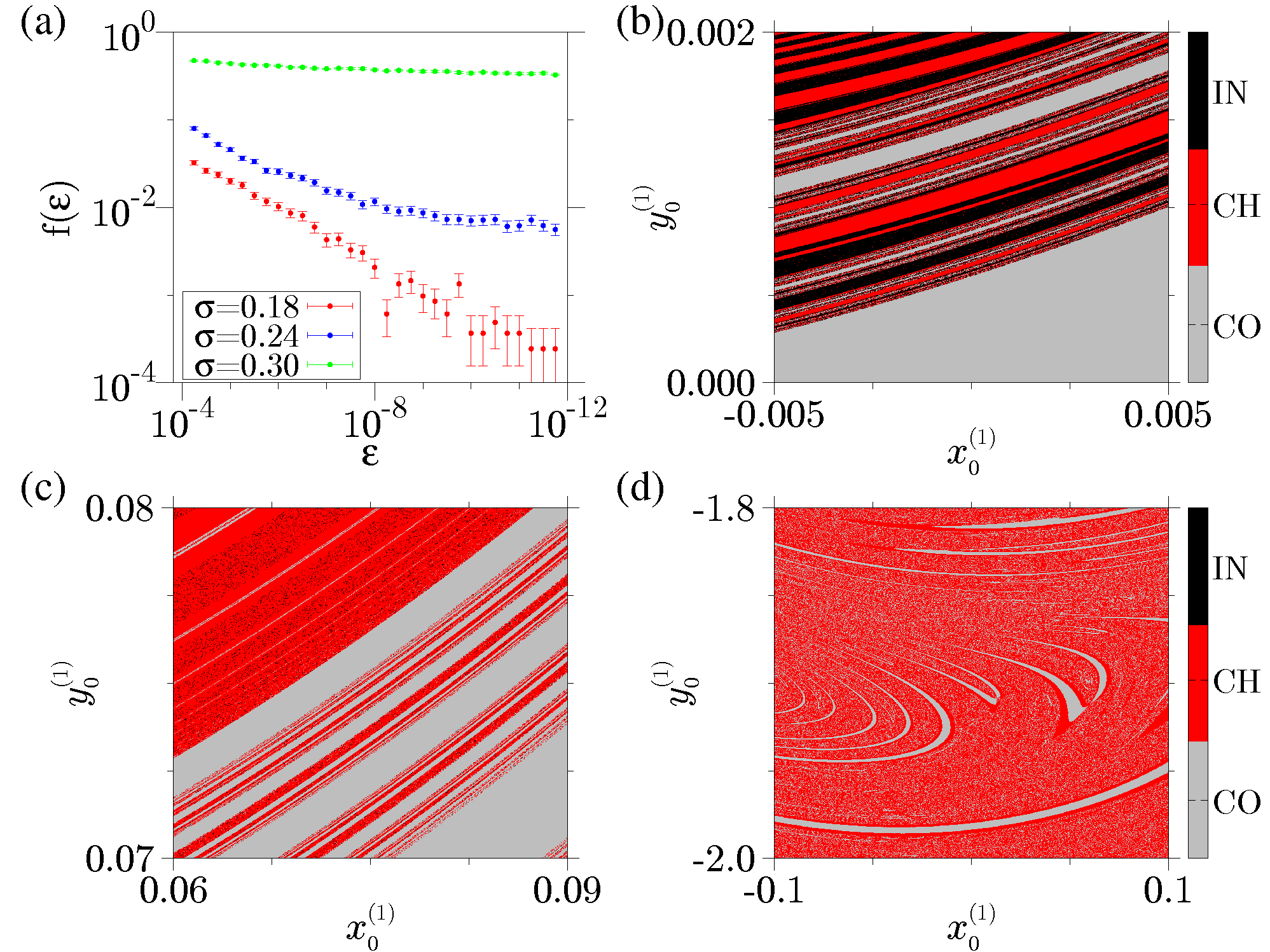}
\caption{(Colour online) (a) Uncertainty fraction $f(\varepsilon)$ versus the
uncertainty radius $\varepsilon$ for the boundary between the chimera and the
coherent basins. Magnification of the basin of attraction for (b)
$\sigma=0.18$, (c) $\sigma=0.24$, and (d) $\sigma=0.30$.}
\label{fig5}
\end{figure}

Firstly, we calculate $f(\varepsilon)$ for the boundary between the chimera and
coherent states basins, as shown in Fig. \ref{fig5}(a). Figures
\ref{fig5}(b)-(d) show magnifications of Figs. \ref{fig4}(b)-(d), that allow to
see the complexity of the boundaries. We use the interval of the magnifications
to estimate $f(\varepsilon)$ versus $\varepsilon$. For $\sigma=0.12$, the basin
of the coherent states is very small, therefore it can be neglected. From the
fitting of the points of Fig. \ref{fig5}(a) we obtain $\gamma=0.30$ (red dots)
$\gamma=0.15$ (blue dots), and $\gamma=0.02$ (green dots) for
$\sigma=0.18, 0.24$, and $0.30$, respectively. As a result, the boundaries
between the chimera and coherent states basins are fractal. A positive and
constant uncertainty coefficient means that the closer you are to an initial
condition, the more likely you are of generating the same final state of the
one generated by that initial condition. The further you go, the more likely
you are changing states by a perturbation in the initial condition. One
consequence of this  observation is that there is a positive probability of a
network in the coherent state to transit to the chimera state if an initial
condition used is perturbed. Since a coherent state can be set by placing all
the initial conditions as equal, it is reasonable to expect that by changing
the initial condition of one node of the network (as we have actually done),
one can reach the chimera state. Another consequence is that the chimera state
can be replaced by the coherent state by a perturbation in the initial
conditions as well. This is a consequence of the fact that the uncertainty
coefficient is positive, and therefore, no matter the precision one alters the
initial conditions, there is always a positive probability for the state to
change. However, since the basin has a fractal boundary, there exist particular
directions to change the initial conditions such that the chimera can be
preserved. This direction is the one associated with the direction where
the dimension is not fractal. All in all, the point is that the chimera state in
the observed network can be found, preserved or altered by design, if one wish
so, as long as the initial conditions are set about the boundary of the
coherent and the chimera states. The same does not happen with respect to the
incoherent state.   

\begin{figure}[hbt]
\centering
\includegraphics[width=0.9\linewidth]{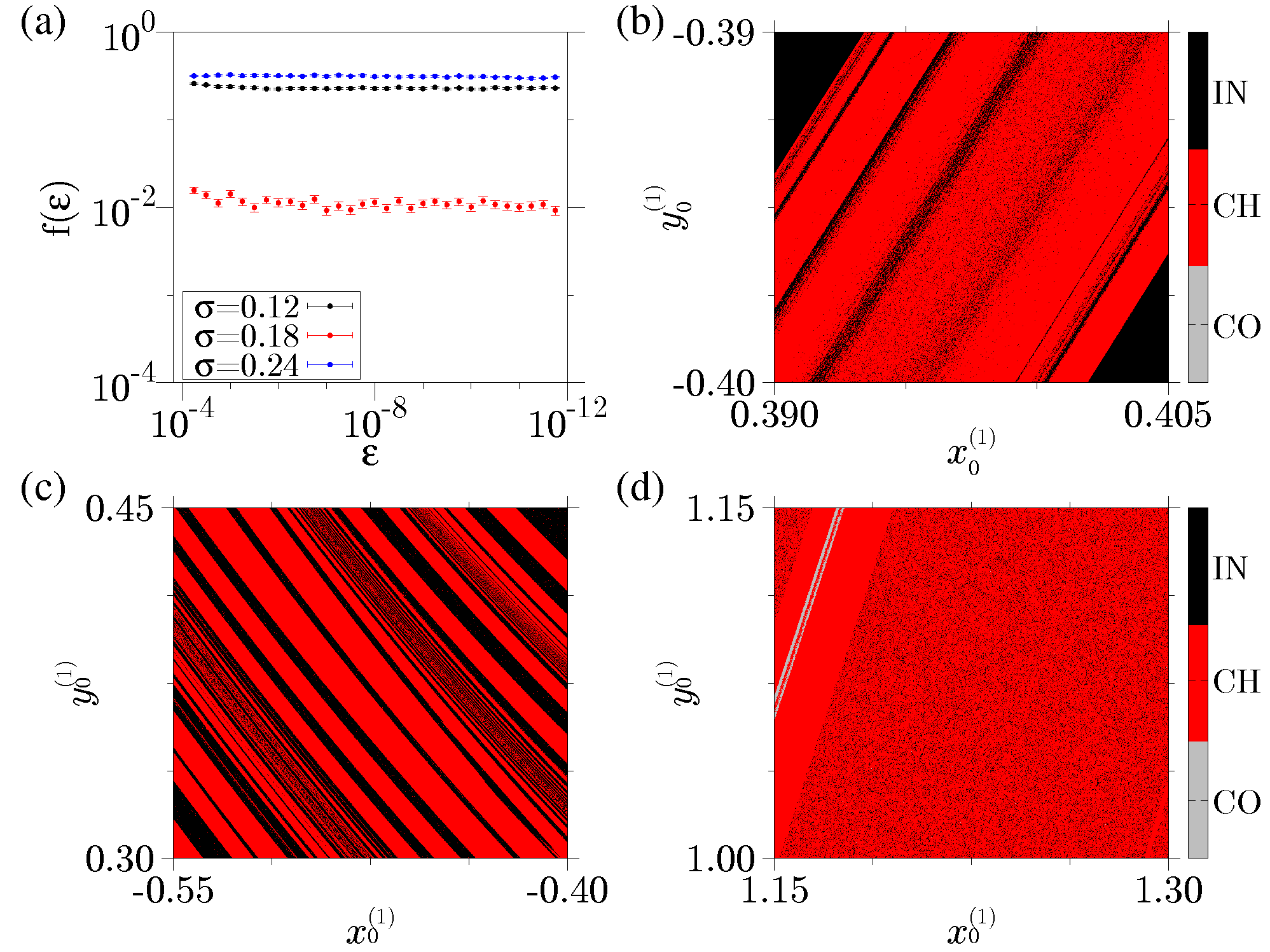}
\caption{(Colour online) (a) Uncertainty fraction $f(\varepsilon)$ versus the
uncertainty radius $\varepsilon$ for the boundary between the incoherent and
chimera states basins. Magnification of the basin of attraction for (b)
$\sigma=0.12$, (c) $\sigma=0.18$, and (d) $\sigma=0.24$.}
\label{fig6}
\end{figure}

Secondly, we compute $f(\varepsilon)$ for the boundary between the chimera and
incoherent states basins, as shown in Fig. \ref{fig6}(a). In Fig.
\ref{fig6}(b)-(d) we plot magnifications of Figs. \ref{fig4}(a)-(c) emphasising
the boundary between incoherent and chimera states basins. The incoherent state
basin has a very small size for $\sigma=0.30$. Our results show that
$f(\varepsilon)$ remains approximately constant for different $\sigma$ values,
and as a consequence $\gamma\approx 0$, indicating the existence of a riddled
basin. A zero uncertainty coefficient means that the probability of finding an
uncertain box, regardless of the resolution of the boxes used (with sides
$\varepsilon$), is constant. No matter how small or large is the perturbation
applied to an initial condition, the change that the system will take place is
the same. This is so because of the riddled basin for which the dimension of
the boundary of the basins of attraction is the dimension of the basin itself.
Thus, in such a situation, it does not exist a special direction for initial
conditions to be perturbed in order to maintain the incoherent state. In
contrast to what was reported before, the preservation or alteration of the
chimera state by a modification on the initial conditions cannot be done by
design, but only in a statistical sense. Therefore, these facts lead us to
conclude that the existence of a riddled basin boundary in a network that
presents chimera is a chimera's dilemma. It makes the state to be fragile by
arbitrarily small changes in the initial conditions.

\section{CONCLUSIONS}

We have analysed a network of circulant coupled H\'enon maps. This network is a
discrete time dynamical system that exhibits coherent and incoherent behaviours.
We consider parameter values where coherent and incoherent domains, named
chimera state, coexist. 

The chimera state coexists with the other two states, namely the coherent and
the incoherent states. All these states have their attraction basin boundaries.
It is known that due to this coexistence, the network may present hysteretic
behaviour as parameters are increased or decreased. The hysteresis character of
the chimera and its coexisting states, where attractors and their basins can
disappear or bifurcate, can potentially provide clarifications about the
emergence of tipping points in nature \cite{medeiros17}. Typically, tipping
points are explained in terms of lower dimensional systems with the coexistence
of states such as equilibrium points or limit cycles. The chimera state could
itself be considered as a possible reason for tipping points emerging in large
dimensional networked systems. Our main interest in this work is to study
properties of the boundary between two of these states, the incoherent and
chimera, and the chimera and the coherent state. Through the uncertainty
exponent, we uncover that the basin boundaries between coherent and chimera
states are fractal, while the basin boundary of incoherent and chimera states
are riddled. Consequently, the first case is more robust to perturbations in
the initial conditions than the second one. Whereas one is likely to obtain a
chimera state by a perturbation of initial conditions leading to the coherent
state (which can be set by having all nodes with the same or roughly the same
initial condition), it is unlikely to appear a chimera state by a perturbation
of initial conditions leading to the incoherent state.


\begin{acknowledgments}
We wish to acknowledge the support: S\~ao Paulo Research Foundation (FAPESP)
under Grants 2011/ 19296-1, 2015/05186-0, 2015/07311-7, 2015/50122-0, and
2017/20920-8, Conselho Nacional de Desenvolvimento Cient\'ifico e Tecnol\'ogico
(CNPq), and Coordena\c c\~ao de Aperfei\c coamento de Pessoal de N\'ivel
Superior (CAPES). 
\end{acknowledgments}

\end{document}